\documentclass{ws-ijmpcs}

\usepackage{units}
\usepackage{hyperref}
\usepackage{textcomp}

\newcommand{\ee}{e^+e^-}
\newcommand{\dzero}{D^0}
\newcommand{\dbarzero}{{}\overline{D}{}^{0}}

\newcommand{\gevcc}{\,\unit{GeV}/\emph{c}^2}

\newcommand{\srd}{r}
\newcommand{\kpi}{{K^-\pi^+}}

\newcommand{\strph}{\delta_{K\pi}}
\newcommand{\mbc}{M_\mathrm{BC}}
\newcommand{\delE}{\Delta\emph{E}}

\begin{document}


%
\catchline{}{}{}{}{}
%

\title{Measurement of strong phase in $D^0{}\to K\pi$ decay at BESIII}

\author{Xiao-Kang ZHOU\\
On Behalf of the BESIII Collaboration}

\address{University of Chinese Academy of Sicences (UCAS)\\
Beijing 100049, China\\
zhouxiaokang08@mails.ucas.ac.cn}

\maketitle


\begin{abstract}
Based on 2.92\,fb$^{-1}$ of $\ee$ collision data collected with the BESIII detector at $\sqrt{s}$ = 3.773~GeV, we measured the asymmetry $\mathcal{A}_{CP\to K\pi}$ of the branching fractions of $D^{CP\pm}\to K^-\pi^+$ ($D^{CP\pm}$ are the $CP$-odd and $CP$-even eigenstates) to be $(12.77\pm1.31^{+0.33}_{-0.31})\%$.
$\mathcal{A}_{CP\to K\pi}$ is used to extract the strong phase difference $\delta_{K\pi}$ between the doubly Cabibbo-suppressed process $\dbarzero\to K^- \pi^+$ and Cabibbo-favored $D^0\to K^- \pi^+$. By taking inputs of other parameters in world measurements, we obtain $\cos\strph = 1.03\pm0.12\pm0.04\pm0.01$.
This is the most accurate result of $\cos\strph$ to date and can improve the world constrains on the mixing parameters and on $\gamma/\phi_3$ in the CKM matrix.
\end{abstract}

\ccode{PACS numbers:}

\section{Introduction}	
The strong phase difference $\strph$ between the doubly Cabibbo-suppressed (DCS) decay $\dbarzero\to K^-\pi^+$ and the corresponding Cabibbo-favored (CF) $\dzero\to K^-\pi^+$ is denoted as
 \begin{equation}
     \frac{\langle K^-\pi^+|\dbarzero\rangle}{\langle K^-\pi^+|\dzero\rangle}=\frac{\overline{A_f}}{A_f}
    = -\srd e^{-i\strph},\label{5} \nonumber
    \end{equation}
which plays an important role in measurements of $\dzero$-$\dbarzero$ mixing parameters 
and helps to improve accuracy of the $\gamma/\phi_3$ angle measurement in CKM matrix.

In the limit of CP conservation, $\strph$ is the same in the final states of $K^-\pi^+$ and $K^+\pi^-$. Here, the notation of $\dzero\to K^-\pi^+$ is used for CF decay and $\dbarzero\to K^-\pi^+$ for DCS, and its charge conjugation mode is always implied to be included.

Using quantum-correlated technique, $\strph$ can be measured in the mass-threshold production process $\ee\to\dzero\dbarzero$~\cite{Cheng:2007uj}~\cite{Xing:1996pn}~\cite{Li:2006wx}~\cite{Gronau:2001nr}.
In this process, the initial $J^{PC}=1^{--}$ requires the $\dzero$ and $\dbarzero$ are in a $CP$-odd  quantum-coherent state. Therefore, at any time, the two $D$ mesons have opposite $CP$-eigenstates until one of them decays. When omitting the high orders, we obtain the $\mathcal{A}_{CP\to K\pi}$ is the asymmetry between $CP$-odd and $CP$-even states decaying to $\kpi$~\cite{Xing:1996pn}~\cite{Li:2006wx}~\cite{Gronau:2001nr}~\cite{CLEO-c2}
\begin{equation}
\!\!\!\!\!
\mathcal{A}_{CP\to K\pi}\equiv\frac{\Gamma_{D^{CP-}\to \kpi}-\Gamma_{D^{CP+}\to \kpi}}{\Gamma_{D^{CP-}\to \kpi}+\Gamma_{D^{CP+}\to \kpi}} \label{12}\nonumber
\end{equation}
and the relationship between $\strph$ and $\mathcal{A}_{CP\to K\pi}$
\begin{equation}
2\srd \cos\strph + y = (1+R_{\mathrm{WS}})\mathcal{A}_{CP\to K\pi},\label{11}\nonumber
\end{equation}
when $y$ is the mixing parameter, $R_{\mathrm{WS}}$ is the decay rate ratio of the wrong sign process $\dbarzero\to\kpi$ and the right sign process $\dzero\to\kpi$. Therefore, with result of $\mathcal{A}_{CP\to K\pi}$ and external inputs of $y$, $r^2$ and $\unit{R_{WS}}$, we can obtain $\cos\delta_{K\pi}$.

CLEO-c used 818\,$\unit{pb^{-1}}$ of $\psi(3770)$ data and used a global fit method to get the $\cos\delta=1.15^{+0.19+0.00}_{-0.17-0.08}$ (with the external parameters) and $\cos\delta=0.81^{+0.22+0.07}_{-0.18-0.05}$ (without the external parameters), where the uncertainties are statistical and systematic\cite{CLEO-c2}.

\section{Determine $\cos\delta_{K\pi}$ in experiment}
To determine these branching fractions, we follow the `DTag' technique firstly introduced by the MARK-III collaboration~\cite{Baltrusaitis:1985iw}. We select $D\to CP$ without reference to the other particle as single tag (ST) events, and reconstruct $D$ to $CP$ and another $D$ to $K\pi$ as double tag (DT) events. The branching fraction for $D$ decays can be obtained from the fraction of DT events in STs with no need of the total number of $D\overline{D}$ events produced.
\begin{table}[ph]
\tbl{$D$ decay modes.}
{\begin{tabular}{@{}cc@{}} \toprule
Type        & Mode  \\\colrule
Flavored    & $K^-\pi^+, K^+\pi^-$  \\
$CP+$         & $K^+K^-, \pi^+\pi^-, K^0_S\pi^0\pi^0, \pi^0\pi^0, \rho^0\pi^0$  \\
$CP-$         & $K^0_S\pi^0, K^0_S\eta, K^0_S\omega$  \\\botrule
\end{tabular} \label{CP_Mode}}
\end{table}

The total dataset analyzed is about 2.92\,fb$^{-1}$. 5 $CP+$ modes and 3 $CP-$ modes are reconstructed from combinations of $\pi^{\pm}, K^{\pm}, \pi^0, \eta$, and $K^0_S$ candidates with $\pi^0\to\gamma\gamma$, $\eta\to\gamma\gamma$, $K^0_S\to\pi^+\pi^-$ and $\omega\to\pi^+\pi^-\pi^0$, as list in Table~\ref{CP_Mode}. Two kinematic variables are defined; the beam constrained mass $\mbc$ and the energy difference $\delE$
\begin{equation}
   \mbc\equiv\sqrt{E^2_0/c^4-|\vec{p}_\mathrm{D}|^2/c^2},\label{25} \nonumber
\end{equation}
\begin{equation}
    \delE\equiv E_\mathrm{D}-E_0,\label{26} \nonumber
\end{equation}
where $\vec{p}_\emph{D}$ and $E_\emph{D}$ are the total momentum and energy of the $D$ candidate, and $E_0$ is the beam energy. $D$ signal candidates produce a peak in $\mbc$ at the $D$ mass and in $\delE$ at zero. To obtain the ST yields, we fit the $\mbc$ distribution with the mode-dependent requirements on $\delE$. We accept only one candidate per mode per event; when multiple candidates are present, we choose the one with the least $|\delE|$.
\begin{figure*}[pt]
\centering
\includegraphics[width=0.95\linewidth]{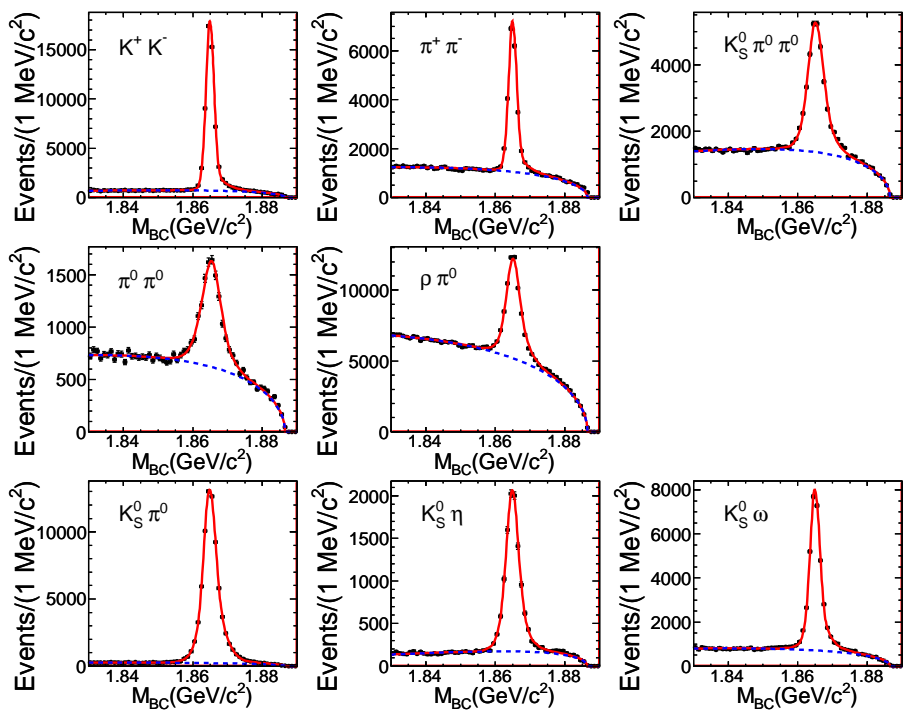}
\caption{ST $\mbc$ distributions of the $D\to CP$ decay final states and fits to the distributions. Data are shown in points with error bars. The solid lines show the total fits and the dashed lines show the background shapes. }{\label{ST_all}}
\centering
\end{figure*}
\begin{figure*}[pt]
\begin{center}
\includegraphics[width=0.9\linewidth]{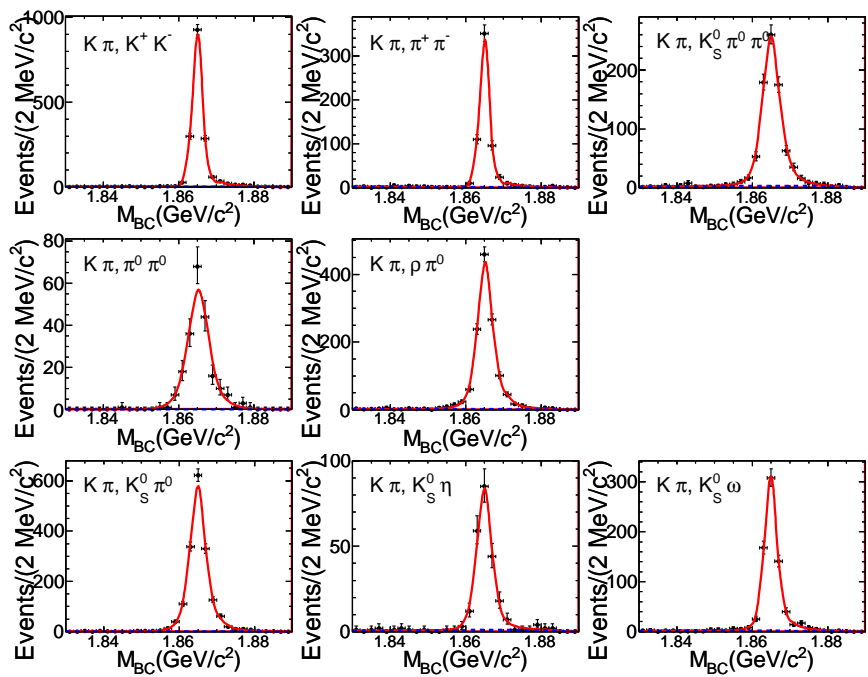}
\caption{DT $\mbc$ distributions and the corresponding fits. Data are shown in points with error bars. The solid lines show the total fits and the dashed lines show the background shapes.}{\label{DT_all}}
\end{center}
\end{figure*}
The $\mbc$ distribution, shown in Figure~\ref{ST_all}, which
has a peak at the mass of the $D$ meson and a smooth background cut-off at the point of the beam energy, are fitted by a signal shape derived from MC simulation, convoluted with a smearing Gaussian function, and a background function modeled with the ARGUS function~\cite{Argus}. The Gaussian function is supposed to compensate for the difference in the signal shape between data and MC simulation.

To select DT signals, we require the $\delE$ of both $D$ mesons, restrict the mass window $1.86\gevcc<\mbc(D\to K\pi)<1.875\gevcc$, and extract the signal yields of DTs by fitting the distributions of $\mbc(D\to CP)$. The $\mbc(D\to CP)$ signal is described by the signal MC shape convoluted with a Gaussian function, and the background is modeled with the ARGUS function.

We obtain $\mathcal{A}_{CP\to K\pi}=(12.77 \pm 1.31^{+0.33}_{-0.31})\%$, where the first uncertainty is statistical and the second is systematic. When quote the external inputs of of $R_\mathrm{D}=3.47\pm 0.06$\textperthousand, $y=6.6\pm0.9$\textperthousand~ from HFAG~\cite{HFAG}, and $R_\mathrm{WS}=3.80\pm0.05$\textperthousand~ from PDG~\cite{PDG}, we obtain $\cos\delta_{K\pi} = 1.03\pm0.12\pm0.04\pm0.01$, where the first uncertainty is statistical, the second uncertainty is systematic, and the third uncertainty is due to the errors introduced by the external inputs.

\section{Summary}
We employ a double tagging technique to analyze a 2.92\,fb$^{-1}$ quantum-correlated data of $\ee\to D^0\dbarzero$ at the $\psi(3770)$ peak. We measure the asymmetry $\mathcal{A}_{CP\to K\pi} = (12.77\pm1.31^{+0.33}_{-0.31})\%$. Using the inputs of $r^2$, $y$ from HFAG and $R_\mathrm{WS}$ from PDG, we get $\cos\delta_{K\pi} = 1.03\pm0.12\pm0.04\pm0.01$. The first uncertainty is statistical, the second is systematic, and the third is due to the external inputs. This is the most precise measurement of the $D\to K\pi$ strong phase difference to date. It helps to constrain the measurements of the $\dzero$-$\dbarzero$ mixing parameters and the $\gamma/\phi_3$ angles in the CKM matrix.


\end{document}